\documentclass[sigconf,review=false]{acmart}
\usepackage{acronym}
\usepackage{balance}
\usepackage{booktabs}
\usepackage{eepic}
\usepackage{hyperref}
\usepackage{graphicx}
\usepackage{framed}
\usepackage[capitalise]{cleveref}
\usepackage{subcaption}

\newcommand{\tablestyle}[0]{\centering\sffamily\small}
  
\acrodef{CWE}{Common Weakness Enumeration}
\acrodef{ISA}{Instruction Set Architecture}
\acrodef{OWASP}{Open Web Application Security Project}
\acrodef{SMB1}{Super Mario Bros{.} 1}
\acrodef{SMB3}{Super Mario Bros{.} 3}
\acrodef{SMW}{Super Mario World}
\acrodef{SM64}{Super Mario 64}
\acrodef{TAS}{Tool Assisted Speedrun}
\acrodef{CWE-252}{Unchecked Return Value}
\acrodef{CWE-367}{Time-of-check Time-of-use (TOCTOU) Race Condition}
\acrodef{CWE-400}{Uncontrolled Resource Consumption}
\acrodef{CWE-459}{Incomplete Cleanup}
\acrodef{CWE-655}{Insufficient Psychological Acceptability}
\acrodef{CWE-696}{Incorrect Behavior Order}
\acrodef{CWE-703}{Improper Check or Handling of Exceptional Conditions}
\acrodef{CWE-755}{Improper Handling of Exceptional Conditions}
\acrodef{CWE-770}{Allocation of Resources Without Limits or Throttling}
\acrodef{7PK}{Seven Pernicious Kingdom}

\copyrightyear{2024}
\acmYear{2024}
\setcopyright{rightsretained}
\acmConference[GAS '24 ]{2024 ACM/IEEE 8th International Workshop on Games and Software Engineering }{April 14, 2024}{Lisbon, Portugal}
\acmBooktitle{2024 ACM/IEEE 8th International Workshop on Games and Software Engineering (GAS '24 ), April 14, 2024, Lisbon, Portugal}\acmDOI{10.1145/3643658.3643917}
\acmISBN{979-8-4007-0561-8/24/04}

\begin{document}
\title{Super Mario in the Pernicious Kingdoms}
\subtitle{Classifying glitches in old games}
\author{Llewellyn Forward}
\affiliation{
%\institution{University of Bristol}
\country{United Kingdom}
}
\authornote{Both authors contributed equally to the paper.}
\author{Io Limmer}
\affiliation{
%\institution{University of Bristol}
\country{United Kingdom}}
\authornotemark[1]
\author{Joseph Hallett}
\affiliation{\institution{University of Bristol}\country{United Kingdom}}
\author{Dan Page}
\affiliation{\institution{University of Bristol}\country{United Kingdom}}
%\acmConference[GAS 2024]{8th International Workshop on Games and Software Engineering}{April 14--20, 2024}{Lisbon Portugal}

\begin{abstract}
  %Super Mario 64\\
  %Has many bugs, and way more than 4.\\
  %SMB three\\
  %As you can plainly see\\
  %Was written in assembly gore!\\
  % ⧣ mods-keyed 'Here is a LaTeX paper.  Could you summarize it in the form of a limerick?' <paper.tex 
% There once was a game, Super Mario,
% With glitches that speedrunners know.
% They exploit the weaknesses,
% Found in the code pieces,
% And beat the game in record time, yo!
% 
%  ⧣ mods-keyed 'Here is a LaTeX paper.  Could you summarize it in the form of a haiku?' <paper.tex 
% Mario glitches found
% Speedrunners exploit their flaws
% Software bugs exposed

In a case study spanning four classic Super Mario games and the analysis of 237known glitches within them, we classify a variety of weaknesses that are exploited by speedrunners to enable them to beat games quickly and in surprising ways. Using the \aclp{7PK} software defect taxonomy and the \acl{CWE}, we categorize the glitches by the weaknesses that enable them.
We identify 7new weaknesses that appear specific to games and which are not covered by current software weakness taxonomies.
%We discuss why games differ from traditional software, and software defect tracking outside of software engineering discipline and inside the speedrunning community.
\end{abstract}
\maketitle

% 1 page
\section{Introduction}

Software has bugs, defects which cause crashes and unwanted behavior and can lead to exploitation. Video games have \emph{glitches}---software \emph{bugs} that cause crashes or unusual behavior, and \emph{sometimes} allow speedrunners (people who compete to beat games in record times) to go fast.
This paper explores the glitches in four classic Super Mario games to see what kinds exist in these games, are if they are the same as the bugs exploited in conventional software.

Categorizing bugs in software allows developers to understand similar problems and bugs.  Various taxonomies have been proposed---from Tsipenyuk~et~al{.}'s \acp{7PK}~\cite{tsipenyuk2005seven}, \acs{OWASP}'s various \emph{top ten} lists~\cite{owasp2021top10} to MITRE's community-developed \ac{CWE}.  Unlike bugs in other software, video game glitches are not always seen as being negative. Instead the weird glitches and \emph{jank}\footnote{A term used to describe glitches in games that are funny or unexpected.} that enable the player to make silly animations or catapult themselves around unpredictably add to the overall aesthetic of the game and make it memorable~\cite{marshall2021janky}.  The speedrunning communities take these glitches further using them to beat games as quickly as possible and raising money for charity~\cite{sher2019speedrunning}. Because the glitches speedrunners exploit may not lead to crashes, little work has examined glitches from a software engineering perspective; yet they have the same root-causes as any other bug---writing correct software is hard.

Nintendo's \emph{Super Mario} is the quintessential video game.  To understand to sorts of glitches speedrunners exploit, we  examine four of earliest Mario platforming games (\ac{SMB1} (1985), \ac{SMB3} (1988), \ac{SMW} (1990) and \ac{SM64} (1996)).  Whilst these games are old, they are still competitively run by speedrunners (see \cref{tab:speedrun.com}) with new records reported in the news~\cite{orland2023record}. The games are well also well understood, having been studied by speedrunners for decades, ensuring that there are large numbers of well researched glitches for analysis.

\begin{table}
  \tablestyle
  \caption{Current world record times for various Any\% categories from \url{speedrun.com}.}
    \begin{tabular}{l l l l l}
      \toprule
      Game & Category & Runner & Time & Date\\
      \midrule
      \acl{SMB1} & Any\%  & Niftski & 4m 54s & 2023-09-06 \\
      \acl{SMB3} & Any\%  & Kuto1k  & 3m 00s & 2023-07-20 \\
      \acl{SMW}  & 0 Exit & FURiOUS & 0m 41s & 2020-06-11 \\
      \acl{SM64} & 0 Star & Suigi   & 6m 20s & 2023-04-16 \\
      \bottomrule
    \end{tabular}
  \label{tab:speedrun.com}
\end{table}

\subsection{Contributions}
This paper addresses the following research question:
\emph{what glitches are speedrunners exploiting, and how do they relate to conventional software bugs?}

This research is important as it lets us understand classes of software bugs (specifically \emph{glitches}) that previous research has overlooked.
Whilst we may not want to fix glitches (they don't always hurt gameplay and can be appreciated by players), they have the same root cause as bugs.  Better understanding how they appear in games helps us better understand how bugs appear in software more generally.

We find evidence for 7new software weaknesses (programming practice that can lead to bugs) that modern software engineering taxonomies (such as \ac{CWE}) have missed.
Whilst some of the glitches speedrunners exploit are similar to conventional software bugs, others seem specific to games and allow speedrunners to beat games fast by exploiting buggy object boundaries or error correction codes.

% 1 page
\section{Background} % The problem

%\subsection{Super Mario}
%\begin{table}
  %\tablestyle
  %\caption{Platforms and release dates for the various Super Mario games studied.}
  %\begin{tabular}{lll}
    %\toprule
    %Game & Platform & Released \\
    %\midrule
    %\acl{SMB1} & NES  & 1985 \\
    %\acl{SMB3} & NES  & 1988 \\
    %\acl{SMW}  & SNES & 1990 \\
    %\acl{SM64} & N64  & 1996 \\
    %\bottomrule
  %\end{tabular}
  %\label{tab:mariogames}
%\end{table}

In the \emph{Super Mario} platforming games
% \footnote{Mario has also been seen go-kart racing, teaching mathematics and typing, partying, playing various sports and featured in two movies.}
 Mario must rescue Princess Peach by jumping through an obstacle course of various platforms to reach a goal, avoiding baddies or defeating them by jumping on their heads.  Players can collect power-ups along the way to unlock special abilities, and coins to increase their score.  The Mario series of games is one of Nintendo's flagship products, and one of the most influential video game series of all time~\cite{loguidice2009vintage}.  

Many early video games, such as the Super Mario games we have examined, were written for consoles that differ from the more uniform PC-like hardware of modern gaming systems~\cite{huang2003xbox}.  Constraints stemming from the hardware, such as limited memory and buses, meant that aggressive optimization and tricks were required to make games run playably~\cite{aycock2016retrogame,montfort2009racing}.  Many of these techniques (for example, the NES's memory mapping~\cite{copetti2022nes}) are niche and known to lead to bugs.  Programming for these systems is closer to embedded development than most modern software, as it requires working around the limits of the hardware to create games.  Despite the challenges of programming these systems, new games are still released~\cite{iam8bit2022garbage} and retro-inspired systems such as PICO-8 inherit some of their architectures.

\Ac{SMB1} and \ac{SMB3} were released for the NES in the 1980s. Each feature eight \emph{worlds} with multiple 2D \emph{levels} that are played in succession.  \Ac{SMW} on the SNES  featured more which could be played multiple times, with multiple exits, as well as new power ups, jump types and enemies.  
Each of these first three games were produced by Shigeru Miyamoto and Takashi Tezuka, and programmed by Toshihiko Nakago (among others).  As such they make for a fascinating case study in the development of glitches in video games: iconic games, created by a fairly consistent team of developers over a five-year period, with each game more complex than before.

\Ac{SM64} is somewhat different to the original Super Mario games.  Created for the N64 it features 3D game play instead of purely 2D platforming.  Instead of linear levels, Mario must traverse large 3D spaces to find stars in which can be collected out-of-order.  The game has complex movement and jump types and introduced many modern game mechanics (such as movable cameras and 360-degree analogue controls) that are now ubiquitous.  
Unlike earlier games, \ac{SM64} was developed by a larger team of programmers and designers, though kept Shigeru Miyamoto as the games producer and director.  In contrast to the first three Super Mario games \ac{SM64} provides a case study for when similar game mechanics (jumping towards a goal) are applied to a wildly different platform and world.  It let us explore if earlier glitches (produced by similar teams), persist when the applied to entirely new game play.

All the games discussed have been released multiple different times by Nintendo, ported to new consoles and systems, and (in the case of \ac{SM64}'s Nintendo DS version) recreated from scratch with new mechanics added.  In this paper we limit ourselves to the releases for the original platforms (though do include glitches from different versions on the same platform).

\begin{figure}
  \tablestyle
\newcommand{\mariosprite}[2]{\subcaptionbox{#1}{\begin{minipage}{38pt}\centering\includegraphics[height=36pt]{#2}\end{minipage}}}
  \mariosprite{Super Mario (SMB1)}{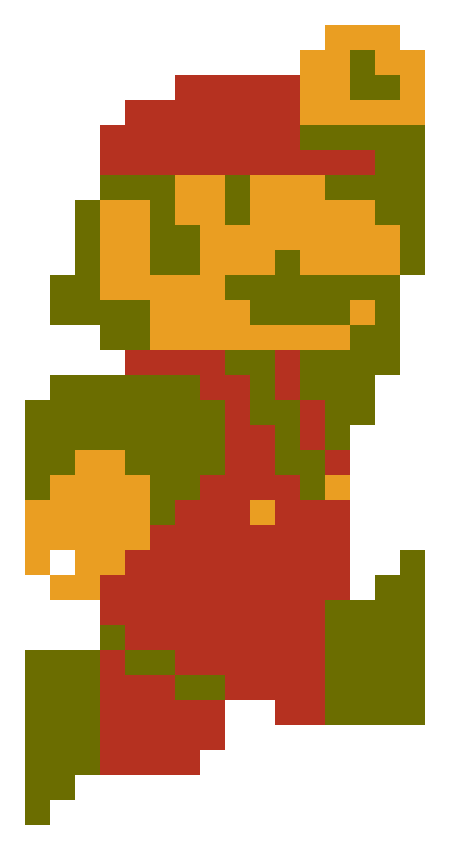}\hfill
  \mariosprite{Bob-omb (SM64)}{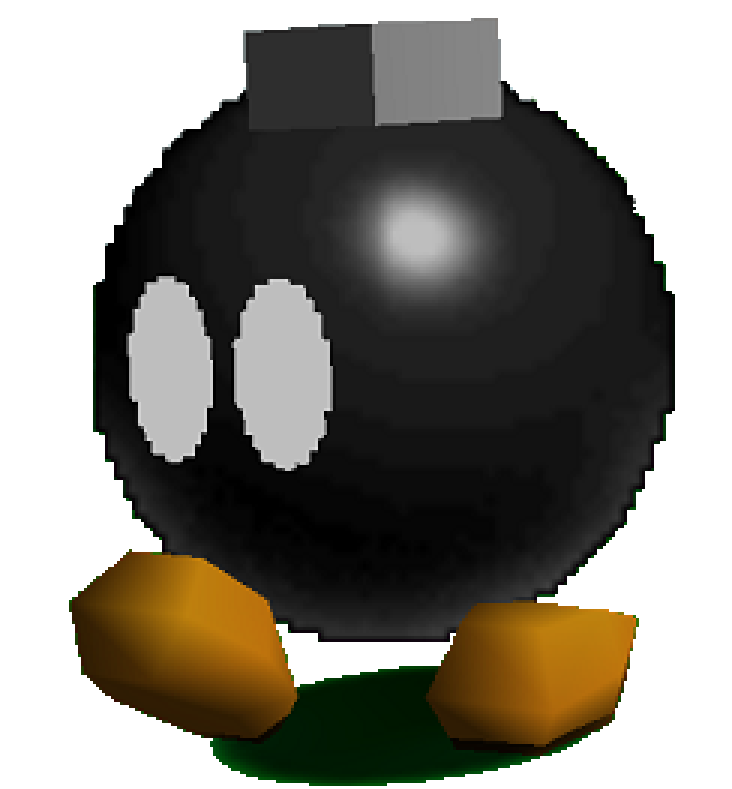}\hfill
  \mariosprite{Bowser (SM64)}{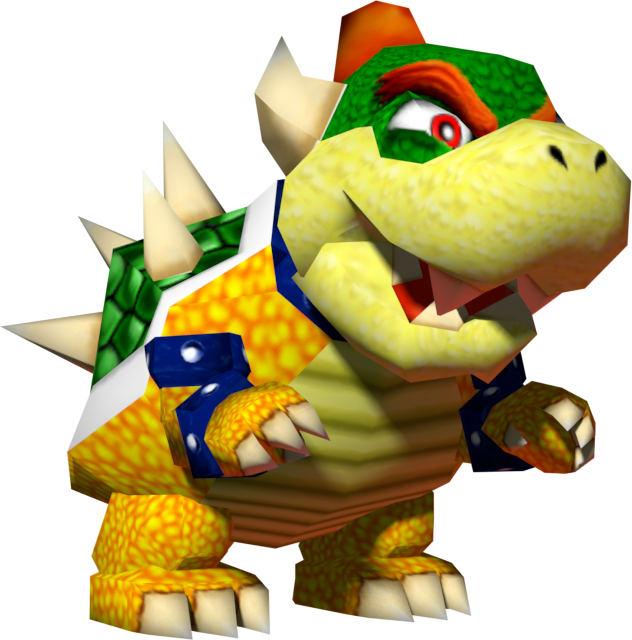}\hfill
  \mariosprite{Chain Chomp (SM64)}{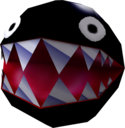}
  \mariosprite{Goomba (SMB3)}{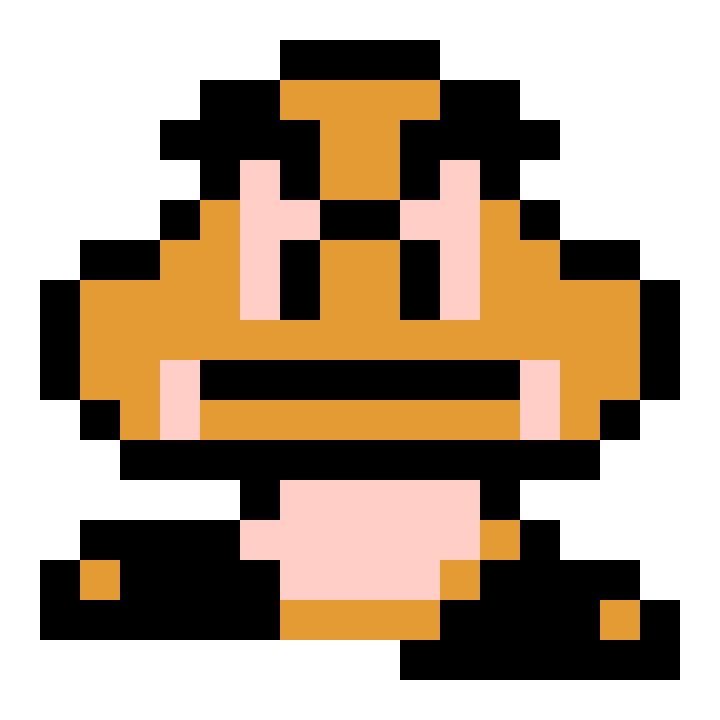}\hfill
  \mariosprite{Koopa Trooper (SMB1)}{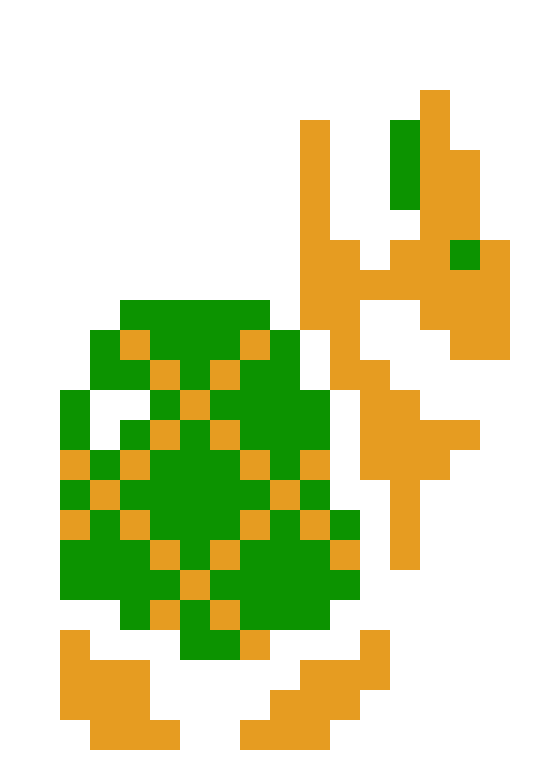}\hfill
  \mariosprite{MIPS the Rabbit (SM64)}{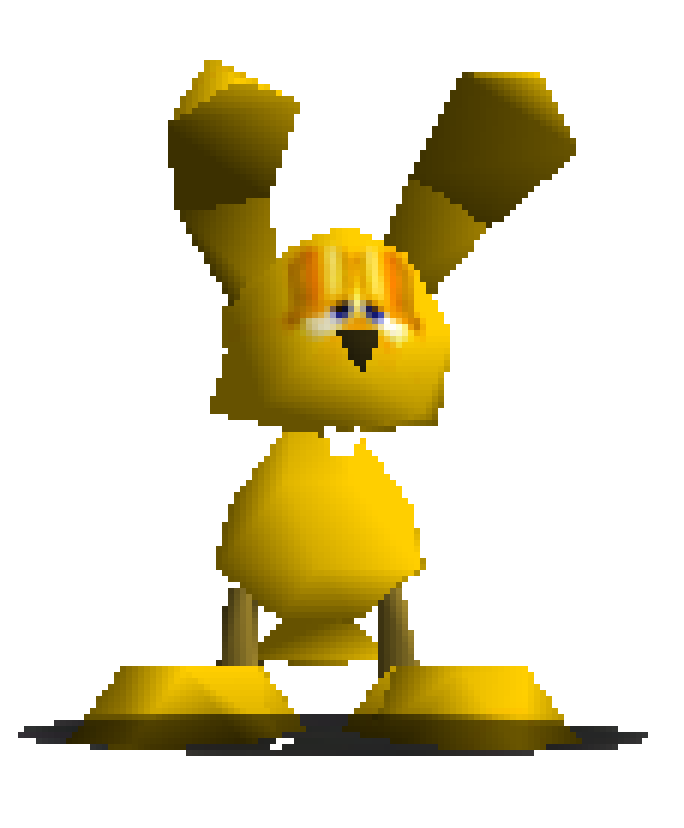}\hfill
  \mariosprite{Bullet Bill (SMW)}{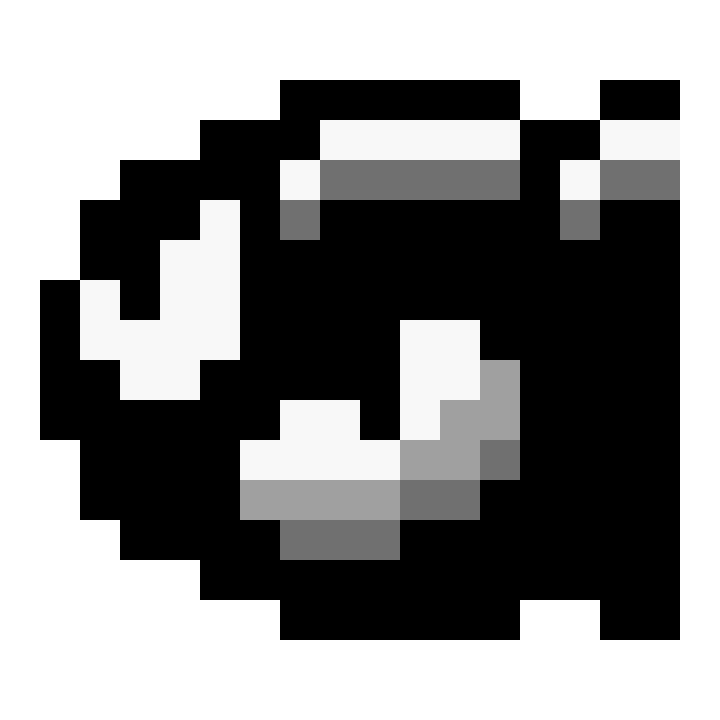}
  \mariosprite{Monty Mole (SM64)}{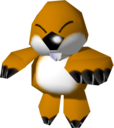}\hfill
  \mariosprite{Thwomp (SMW)}{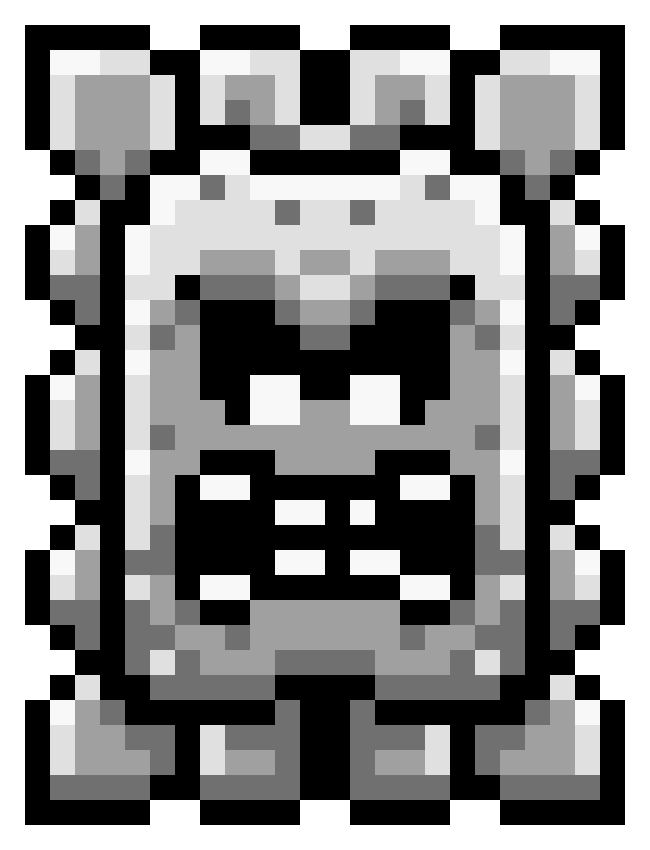}\hfill
  \mariosprite{Piranha Plant (SMW)}{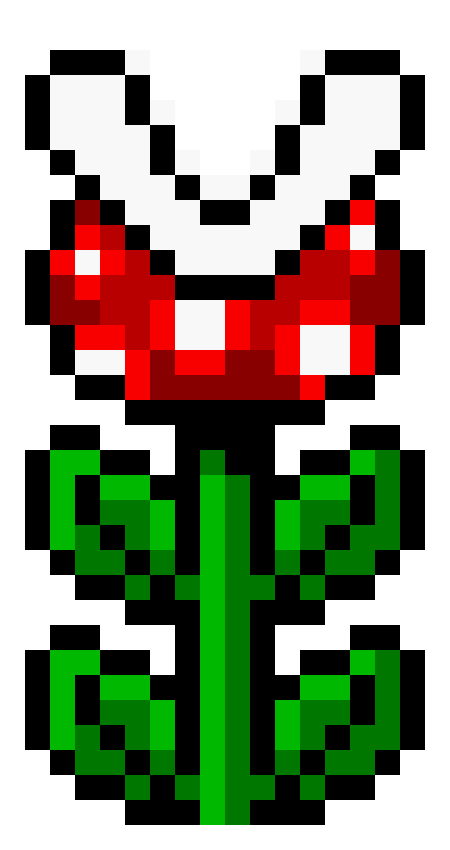}\hfill
  \mariosprite{Boo (SMW)}{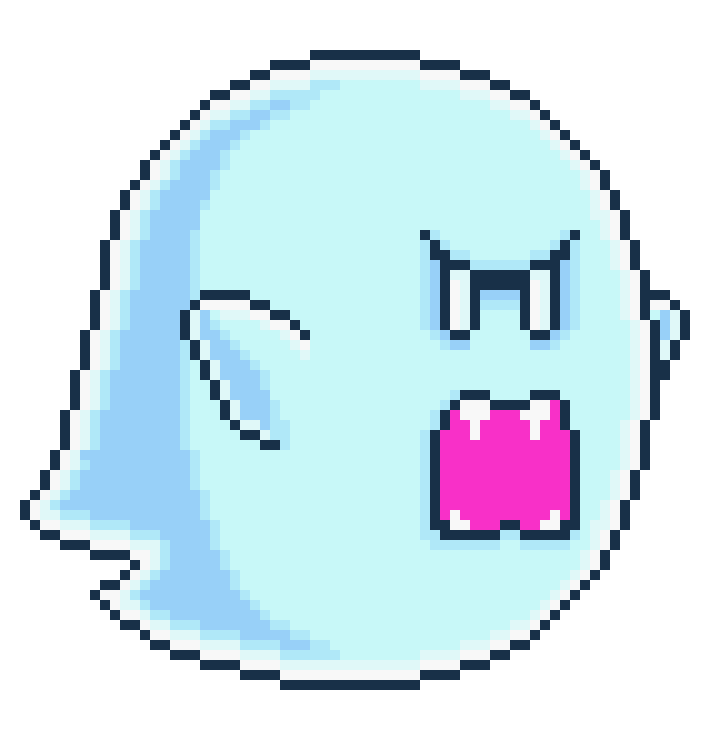}
  \caption{The Mario games have featured many different characters and items which may or may not be familiar. We show a selection from various games to jog the reader's memory.}
  \label{fig:sprites}
\end{figure}

\subsection{Speedrunning and glitch hunting}

Speedrunning is
\emph{``the practice of players or \emph{runners} attempting to \emph{travel} from a game's opening state at its first necessary button input to the game's conclusion at its last necessary button input in the smallest amount of time possible.''}~\cite{scullyblaker2014practiced}

Leaderboards for many games are available online\footnote{In particular \url{https://speedrun.com/}.} and runners compete to beat various games in record times in various categories and upload videos of their \emph{runs} to prove their mastery of the game.  Unlike retro software more generally, older games are still actively and competively run (\cref{tab:speedrun.com}), with new glitches being actively sought\footnote{We speculate that there are very few people still searching for bugs in \emph{Lotus 1--2--3}.}---making the lifespan of retro games somewhat unique.

Alongside 100\% runs (where the player must complete the game fully) and the glitchless categories (where the player must complete the game largely as the developers intended) are the \emph{Any\%} categories where a player can get to the ending by any means necessary---including through glitches and the  exploitation of bugs in the game's programming. 

An example of this is \emph{SethBling}'s 45s run of \ac{SMW}\footnote{Explained by SethBling on YouTube at: \url{https://youtu.be/Jf9i7MjViCE}.}. To beat the game in under a minute they manipulate the position of sprites pixel perfectly on the first level, by jumping and throwing shells, to ensure a pointer is present in memory before abusing a heap overflow-esque vulnerability in memory handling of how sprites are loaded to execute a \emph{jump to subroutine} instruction with their injected pointer.  The subroutine they load is the end of game credit sequence, thus completing the game without ever completing a single level.  Similar bugs were later employed to dynamically reprogram the game to be Nintendo's 2015 \emph{Super Mario Maker} for a \ac{TAS}\footnote{A speedrunning category where the inputs to the game can be preprogrammed and replayed for levels of input precision far greater than a human is capable of.} demonstration at a speedrunning event~\cite{masterjun2016tasbot}.

In conventional software, the style and the mechanisms used by Any\% and \ac{TAS} speedrunners are similar to offensive cybersecurity techniques, albeit with different goals.  The engineering and speedrunning communities remain separate.  Bug discovery techniques, such as fuzzing or symbolic execution, are not typically deployed as part of glitch discovery in games.  Instead glitch hunters appear to work through reverse engineering, attempting to recreate isolated reports of the glitches (for example a speedrunner was able to complete a level in \ac{SM64} through what is believed to have been a freak cosmic ray flipping a bit in the game's memory~\cite{day2021cosmic}), and in some cases by exhaustively trying the same actions repeatedly \emph{in case something goes wrong} (for example Nintendo's The Legend of Zelda Wind Waker's \emph{Barrier Skip} which was speculated as being possible, but took years to be performed accidentally and further more years to be reproduced~\cite{alexandra2019windwaker}).

\subsection{Pernicious Kingdoms}
\label{sec:pernicious-kingdoms}

As part of more general software bug tracking, MITRE's \ac{CWE} taxonomy attempts to describe the underlying weaknesses that enabled the bugs to occur in the first place.  To further categorize the bugs various attempts have been made to impose some further structure on the \ac{CWE} list and rank the \emph{importance} of each class of bug, as well as to create sub-taxonomies for particular communities (for example CWE-660 captures relationships to other \acp{CWE} that may be specific to Java programming).

Tsipenyuk~et~al{.}'s \acfp{7PK}~\cite{tsipenyuk2005seven} is an attempt to provide a high-level categorization of software bugs---and inspiration for the CWE taxonomy.  A mapping of several \acp{CWE} to the \acp{7PK} is given in CWE-700. The \acp{7PK} describe eight\footnote{The authors believe that humans can process information in groups of $7\pm{2}$, so opted to call the paper the \emph{seven} pernicious kingdoms despite having \emph{eight} of them.} categories of vulnerabilities each described roughly in order of its importance (\cref{tab:7pk}).  Many other general taxonomies (for example, OWASP's Top 10~\cite{owasp2021top10} and the 19 and subsequent 24 Deadly Sins of Software Security~\cite{blanc2005deadly,howard2010deadly}) exist but \ac{7PK} claims to improve over them by simplifying more.  We use the \acp{7PK} as higher level categories for discussing the glitches we find in the games examined.

\begin{table}
  \tablestyle
  \caption{The Seven Pernicious Kingdoms (from~\cite{tsipenyuk2005seven}).}
  \begin{tabular}{
    p{\dimexpr 0.37\linewidth - 2\tabcolsep}
    p{\dimexpr 0.63\linewidth - 2\tabcolsep}
  }
    \toprule
    Kingdom & Description \\
    \midrule
7PK1: Input Validation and Representation  & Problems caused by trusting input, including validation and representation issues.  This kingdom includes buffer overflows. \\
7PK2: API Abuse  & Issues caused by not using an API correctly and violating API assumptions. \\
7PK3: Security Features  & Cryptography and access control issues. \\
7PK4: Time and State  & Concurrency issues and problems caused by the order in which an object's state is manipulated. \\
7PK5: Errors  & Similar to API abuse but specialized around how programmers handle exceptions. \\
7PK6: Code Quality  & Unpredictable behavior from dodgy code. \\
7PK7: Encapsulation  & Errors arising from how different software components interact. \\
7PK*: Environment  & Errors arising from how software interacts with everything that is not software (for example: hardware, configuration files\ldots).\\
    \bottomrule
  \end{tabular}
  \label{tab:7pk}
\end{table}

Other papers have tried to classify the causes of glitches in games.  Bainbridge and Bainbridge categorized 751 glitches in 155 games and made an attempt to assign a cause to 580 of the glitches~\cite{bainbridge2007creative}.   They created a hierarchy of ten codes organized into three categories to capture the underlying cause.  Their codes capture high level causes for the glitches, but do not explore what flaws in the software cause the glitches.  For example, Bainbridge and Bainbridge have a category \emph{bad code} to capture glitches cased by generic programming errors. We map instead at the \ac{CWE} level identifying what bad coding pattern enabled the weakness to begin with.
Lewis~et~al{.}  proposed a taxonomy of bugs in video games, though did not attempt to speculate on the causes of the bugs~\cite{lewis2010what}.   They split their issues into \emph{temporal} and \emph{non-temporal} issues categorising how the glitches displayed rather than what went wrong.

%\begin{figure}
%  \tablestyle
%  \includegraphics[width=0.8\linewidth]{figures/out/bainbridge.pdf}
%  \caption{Bainbridge and Bainbridge's categorization of 580 glitches~\cite{bainbridge2007creative}.}
%  \label{fig:bainbridge}
%  \end{figure}
%
%\begin{figure}
%  \tablestyle
%  \includegraphics[width=1.0\linewidth]{figures/out/lewis.pdf}
%  \caption{Lewis~et~al{.}'s taxonomy of bugs in games~\cite{lewis2010what}.}
%  \label{fig:lewis}
%  \end{figure}

\section{Method} % My idea

This \emph{historical case study} was performed following the ACM Empirical Guidelines for Software Engineering Research~\cite{ralph2020empirical}.  From existing lists of glitches for the four games we performed \emph{thematic analysis}~\cite{braun2006using} using an integrated approach~\cite{cruzes2011recommended} with the \ac{CWE} taxonomy~\cite{mitre2006cwe} used as an initial codebook. \Ac{CWE} is designed to be a baseline for weakness identification in software projects and so forms an ideal initial codebook when attempting to identify weaknesses in Super Mario's games.

Initial coding of weaknesses was performed independently by the first and second authors (coding \ac{SMB1}, \ac{SMB3} and \ac{SMW}; and \ac{SM64} respectively).  Triangulation was performed initially independently in discussion with the third author who subsequently merged the first two authors' codebooks and analysis for their games.  The new codebook was discussed and modified until consensus was reached amongst all authors.

\subsection{Data source}

Data for the glitches for each game was taken from the \emph{Super Mario Wiki}\footnote{\url{https://www.mariowiki.com}}.  The wiki collects together various glitches for the Mario games and cross-references them against evidence for the glitch's existence (typically demonstrations by the discoverer on YouTube). Since the glitch hunting and speedrunning communities are themselves informal, unlike the CVE databases, documentation of glitches is largely informal limited to word-of-mouth and private chat channels.  Collected references, such as Super Mario Wiki make a valuable resource for researchers.

To ensure the descriptions of the glitches were accurate, we reproduced all bar three of the glitches for \ac{SMB1}, \ac{SMB3} and \ac{SMW}.  Due to a lack of skill we were unable to reproduce the glitches for \ac{SM64}, though there is greater documentation of those glitches on YouTube.  Our analysis is based on observation of the glitch's effect, and expert analysis rather than an inspection of the game's source code or internal memory state.  We examined 237glitches for four games, mapping them to 32codes.

%\begin{table}
  %\tablestyle
  %\caption{Counts of glitches for each of the games that were analyzed. Each glitch was mapped to a single \ac{CWE}, but \ac{CWE} were mapped to multiple games.}
  %\input{data/out/glitch-counts}
  %\label{tab:glitch-counts}
%\end{table}

\subsection{Mapping process}

To illustrate the mapping process used, consider the \emph{wall jump} glitch from \ac{SMB1}.  It is described as:
\begin{quote}
  ``If Mario's foot catches on a wall or pipe he can jump again to do the Wall Jump at that frame.  However, Mario has to be moving towards the wall with some velocity.  This is because when Mario hits a wall, he goes slightly into the wall.  The bricks in the wall count as individual surfaces, so Mario has a surface to jump off.''---Super Mario Wiki
\end{quote}
On every frame Mario's position is updated on using his current velocity.  This update is done without considering whether a collision has occurred. Instead Mario's position will be corrected (by moving him backwards if he has entered an impossible state) once tthe motion is complete.
Similarly when the player presses the button to make Mario jump, the game runs through a series of tests.  It checks whether there is a block immediately bellow him, and if so starts his jump. What happens if these two actions overlap?  Mario is allowed an extra jump off of the wall, as the jump logic completes before the motion logic.
What is the underlying reason for this glitch? Since the cause is that the jump-logic is completed before the movement-logic, when it might be more normal to do it the other way round we assigned it to \ac{CWE-696}.  This coding process was repeated iteratively for all the glitches examined until consensus was agreed.

%\begin{figure}
  %\tablestyle
  %\includegraphics[width=0.7\linewidth]{figures/out/wall-jump.pdf}
  %\caption{Overlapping jump and movement actions.  Rectangular blocks express movement-logic, ovals express the jump-logic.  The expected sequence of actions is shown by the bold arrow, but the actual behavior is shown by the dotted line.}
  %\label{fig:wall-jump-logic}
%\end{figure}

\subsection{Threats to validity}
We have identified the following threats to the validity of this work:
firstly the glitches we analyzed came from online lists.  They may not be exhaustive nor accurate, and represent a snapshot of the bugs known (and considered important enough to document) by the glitch-hunting communities.  Other bugs may exist.
  
Secondly, the reports of the glitches not be accurate.  For most of the bugs we have either reproduced them ourselves, or relied upon demonstrations of the glitches being performed by others (in particular \emph{Pannenkoek2012}'s series on YouTube for \ac{SM64}~\cite{koek2016monty,koek2016rolling}).
  
Thirdly, we do not have the source code for the games and cannot know if the glitches are intended behavior.  Whilst source code for the games has been reverse engineered and leaked online we have chosen instead to rely on our own knowledge of software engineering when ascribing a root cause for the bug.  It is likely that our coding is influenced by our own experiences.
  Our analysis of the glitches is based on how the glitches appear when playing, rather than an inspection of the internal state of the consoles.  Because many of the games were programmed \emph{close to the metal} making frequent use of assembly and obscure hardware features, our analysis will miss some implementation specific details by focusing on effects of the glitches rather than the hardware producing them.
  
Finally, whilst we allowed for the use of multiple codes per glitch, in practice we almost always only used one.  An example of this is the \emph{Pipe Death} glitch described in \cref{sssec:7pk5}: initially we coded it as \acf{CWE-703} and \emph{Correct But Surprising Behavior}, but on review seeing only a handful of glitches with multiple codes we reduced it to just \emph{Correct But Surprising} as we felt it described the bug sufficiently. 

% 5 pages
\section{Mario in the Pernicious Kingdoms}
\begin{table*}
  \tablestyle
  \caption{Count of CWEs assigned to glitches in each game. Glitches are split by game and also by which of the eight \acp{7PK} it is categorized by.  No glitches were mapped to Pernicious Kingdoms 3 and 7. Absolute and percentage counts are given for each game giving absolute counts of glitches and the total percentage of glitches mapped to each CWE per game and overall.}
  \begin{tabular}{llr@{}rr@{}rr@{}rr@{}rr@{}r}
  \toprule
  & Glitch & \multicolumn{2}{c}{\ac{SMB1}} & \multicolumn{2}{c}{\ac{SMB3}} & \multicolumn{2}{c}{\ac{SMW}} & \multicolumn{2}{c}{\ac{SM64}} & \multicolumn{2}{c}{Total} \\
  \midrule
  7PK1 & CWE-20: Improper Input Validation & 4 & {\scriptsize (12\%)} & 4 & {\scriptsize (7\%)} & 1 & {\scriptsize (3\%)} & 0 & & 9 & {\scriptsize (4\%)} \\
  7PK1 & CWE-120: Buffer Copy without Checking Size of Input & 0 & & 1 & {\scriptsize (2\%)} & 1 & {\scriptsize (3\%)} & 0 & & 2 & {\scriptsize (1\%)} \\
  7PK1 & CWE-125: Out-of-bounds Read & 2 & {\scriptsize (6\%)} & 4 & {\scriptsize (7\%)} & 0 & & 0 & & 6 & {\scriptsize (3\%)} \\
  7PK1 & CWE-190: Integer Overflow or Wraparound & 3 & {\scriptsize (9\%)} & 2 & {\scriptsize (4\%)} & 0 & & 2 & {\scriptsize (2\%)} & 7 & {\scriptsize (3\%)} \\
  7PK1 & CWE-191: Integer Underflow (Wrap or Wraparound) & 0 & & 0 & & 0 & & 1 & {\scriptsize (1\%)} & 1 & {\scriptsize (0\%)} \\
  7PK1 & CWE-192: Integer Coercion Error & 1 & {\scriptsize (3\%)} & 0 & & 0 & & 0 & & 1 & {\scriptsize (0\%)} \\
  7PK1 & CWE-193: Off-by-one Error & 2 & {\scriptsize (6\%)} & 1 & {\scriptsize (2\%)} & 0 & & 0 & & 3 & {\scriptsize (1\%)} \\
  7PK1 & CWE-788: Access of Memory Location After End of Buffer & 1 & {\scriptsize (3\%)} & 1 & {\scriptsize (2\%)} & 0 & & 0 & & 2 & {\scriptsize (1\%)} \\
  7PK1 & CWE-839: Numeric Range Comparison Without Minimum Check & 0 & & 0 & & 0 & & 2 & {\scriptsize (2\%)} & 2 & {\scriptsize (1\%)} \\
  7PK1 & NEW: Entity Limit & 3 & {\scriptsize (9\%)} & 2 & {\scriptsize (4\%)} & 3 & {\scriptsize (8\%)} & 1 & {\scriptsize (1\%)} & 9 & {\scriptsize (4\%)} \\
%\addlinespace
  \multicolumn{2}{l}{\emph{7PK1 Total}} &  16 & {\scriptsize (50\%)} & 15 & {\scriptsize (28\%)} & 5 & {\scriptsize (14\%)} & 6 & {\scriptsize (5\%)} & 42 & {\scriptsize (18\%)} \\
\midrule
  7PK2 & CWE-252: Unchecked Return Value & 0 & & 0 & & 0 & & 1 & {\scriptsize (1\%)} & 1 & {\scriptsize (0\%)} \\
%\addlinespace
  \multicolumn{2}{l}{\emph{7PK2 Total}} &  0 & & 0 & & 0 & & 1 & {\scriptsize (1\%)} & 1 & {\scriptsize (0\%)} \\
\midrule
  7PK4 & CWE-367: Time-of-check Time-of-use (TOCTOU) Race Condition & 0 & & 0 & & 0 & & 17 & {\scriptsize (15\%)} & 17 & {\scriptsize (7\%)} \\
  7PK4 & CWE-372: Incomplete Internal State Distinction & 0 & & 0 & & 0 & & 7 & {\scriptsize (6\%)} & 7 & {\scriptsize (3\%)} \\
  7PK4 & CWE-456: Missing Initialization of a Variable & 0 & & 0 & & 0 & & 5 & {\scriptsize (4\%)} & 5 & {\scriptsize (2\%)} \\
  7PK4 & CWE-459: Incomplete Cleanup & 2 & {\scriptsize (6\%)} & 7 & {\scriptsize (13\%)} & 8 & {\scriptsize (22\%)} & 15 & {\scriptsize (13\%)} & 32 & {\scriptsize (14\%)} \\
  7PK4 & CWE-665: Improper Initialization & 3 & {\scriptsize (9\%)} & 3 & {\scriptsize (6\%)} & 1 & {\scriptsize (3\%)} & 1 & {\scriptsize (1\%)} & 8 & {\scriptsize (3\%)} \\
  7PK4 & CWE-696: Incorrect Behavior Order & 2 & {\scriptsize (6\%)} & 2 & {\scriptsize (4\%)} & 1 & {\scriptsize (3\%)} & 7 & {\scriptsize (6\%)} & 12 & {\scriptsize (5\%)} \\
  7PK4 & CWE-820: Missing Synchronization & 2 & {\scriptsize (6\%)} & 3 & {\scriptsize (6\%)} & 2 & {\scriptsize (6\%)} & 0 & & 7 & {\scriptsize (3\%)} \\
  7PK4 & CWE-821: Incorrect Synchronization & 0 & & 6 & {\scriptsize (11\%)} & 1 & {\scriptsize (3\%)} & 0 & & 7 & {\scriptsize (3\%)} \\
  7PK4 & CWE-833: Deadlock & 0 & & 0 & & 0 & & 3 & {\scriptsize (3\%)} & 3 & {\scriptsize (1\%)} \\
  7PK4 & CWE-908: Use of Uninitialized Resource & 0 & & 0 & & 0 & & 2 & {\scriptsize (2\%)} & 2 & {\scriptsize (1\%)} \\
  7PK4 & NEW: Failure to Appropriately Recheck Condition & 0 & & 1 & {\scriptsize (2\%)} & 1 & {\scriptsize (3\%)} & 0 & & 2 & {\scriptsize (1\%)} \\
  7PK4 & NEW: State Exit Too Early & 0 & & 0 & & 0 & & 5 & {\scriptsize (4\%)} & 5 & {\scriptsize (2\%)} \\
%\addlinespace
  \multicolumn{2}{l}{\emph{7PK4 Total}} &  9 & {\scriptsize (28\%)} & 22 & {\scriptsize (41\%)} & 14 & {\scriptsize (39\%)} & 62 & {\scriptsize (54\%)} & 107 & {\scriptsize (45\%)} \\
\midrule
  7PK5 & CWE-703: Improper Check or Handling of Exceptional Conditions & 2 & {\scriptsize (6\%)} & 8 & {\scriptsize (15\%)} & 4 & {\scriptsize (11\%)} & 15 & {\scriptsize (13\%)} & 29 & {\scriptsize (12\%)} \\
  7PK5 & NEW: Correct But Surprising Behavior & 0 & & 3 & {\scriptsize (6\%)} & 1 & {\scriptsize (3\%)} & 10 & {\scriptsize (9\%)} & 14 & {\scriptsize (6\%)} \\
  7PK5 & NEW: Improper Correction of Illegal State & 2 & {\scriptsize (6\%)} & 0 & & 1 & {\scriptsize (3\%)} & 0 & & 3 & {\scriptsize (1\%)} \\
%\addlinespace
  \multicolumn{2}{l}{\emph{7PK5 Total}} &  4 & {\scriptsize (12\%)} & 11 & {\scriptsize (20\%)} & 6 & {\scriptsize (17\%)} & 25 & {\scriptsize (22\%)} & 46 & {\scriptsize (19\%)} \\
\midrule
  7PK6 & CWE-475: Undefined Behavior for Input to API & 0 & & 1 & {\scriptsize (2\%)} & 2 & {\scriptsize (6\%)} & 0 & & 3 & {\scriptsize (1\%)} \\
  7PK6 & CWE-835: Loop with Unreachable Exit Condition & 0 & & 0 & & 2 & {\scriptsize (6\%)} & 2 & {\scriptsize (2\%)} & 4 & {\scriptsize (2\%)} \\
  7PK6 & CWE-837: Improper Enforcement of a Single, Unique Action & 1 & {\scriptsize (3\%)} & 0 & & 4 & {\scriptsize (11\%)} & 0 & & 5 & {\scriptsize (2\%)} \\
  7PK6 & CWE-1023: Incomplete Comparison with Missing Factors & 0 & & 1 & {\scriptsize (2\%)} & 0 & & 3 & {\scriptsize (3\%)} & 4 & {\scriptsize (2\%)} \\
  7PK6 & NEW: Incorrectly Shared Behavior & 1 & {\scriptsize (3\%)} & 3 & {\scriptsize (6\%)} & 1 & {\scriptsize (3\%)} & 0 & & 5 & {\scriptsize (2\%)} \\
%\addlinespace
  \multicolumn{2}{l}{\emph{7PK6 Total}} &  2 & {\scriptsize (6\%)} & 5 & {\scriptsize (9\%)} & 9 & {\scriptsize (25\%)} & 5 & {\scriptsize (4\%)} & 21 & {\scriptsize (9\%)} \\
\midrule
  7PK* & NEW: Bad Boundary Definition & 1 & {\scriptsize (3\%)} & 1 & {\scriptsize (2\%)} & 2 & {\scriptsize (6\%)} & 16 & {\scriptsize (14\%)} & 20 & {\scriptsize (8\%)} \\
%\addlinespace
  \multicolumn{2}{l}{\emph{7PK* Total}} &  1 & {\scriptsize (3\%)} & 1 & {\scriptsize (2\%)} & 2 & {\scriptsize (6\%)} & 16 & {\scriptsize (14\%)} & 20 & {\scriptsize (8\%)} \\
\midrule
  \multicolumn{2}{l}{\emph{Grand total glitch count}} & \multicolumn{2}{c}{32} & \multicolumn{2}{c}{54} & \multicolumn{2}{c}{36} & \multicolumn{2}{c}{115} & \multicolumn{2}{c}{237} \\
  \bottomrule
\end{tabular}

  \label{tab:codes-count}
\end{table*}
\begin{table*}
  \caption{Codebook of new categories of weakness for games that did not seem to be part of \ac{CWE}.}
  \tablestyle
  \begin{tabular}{p{\dimexpr 0.3\linewidth-2\tabcolsep} p{\dimexpr 0.7\linewidth-2\tabcolsep}}
    \toprule
    Weakness & Description \\
    \midrule
    Entity Limit & The game attempts to load elements when there is only space for a finite number of them.\\
    Failure to Appropriately Recheck Condition & The game fails to recheck a condition when two objects interact leading to odd behavior. \\
    State Exit Too Early & The game allows an object to transition out of a state change when it should be allowed to complete. \\
    Correct But Surprising Behavior & The game's behavior is surprising, but consistent with its intended programming. \\
    Improper Correction of Illegal State & In attempting to fix an exception, the game has created a new error. \\
    Incorrectly Shared Behavior & Two objects share behavior when they should behave differently. \\
    Bad Boundary Definition & The boundaries of an object in the game are improperly specified. \\
    \bottomrule 
  \end{tabular}
  \label{tab:codebook}
\end{table*}
\Cref{tab:codes-count} shows the breakdown of glitches mapped per game.  For each \ac{CWE} we group them by the \acp{7PK}, using the mapping relationship from \ac{CWE} or its parent \ac{CWE}; and using our own judgment for those without mappings.  Using the \acp{7PK} as high level groups we discuss what glitches exist in order of prevalence.

\subsection{7PK4: Time and State}

45\% of the glitches could be attributed to the \emph{Time and State} kingdom.  Video games are, in essence, giant state machines with intricate interactions between elements: sometimes these interactions get confused and glitches appear.
Within the kingdom, the largest source of glitches in general was \ac{CWE-459}.  This appeared as an \emph{invisible vine} in \ac{SMB1}---where if Mario died from a hammer in level 5--2 while climbing a vine, then on restarting the level Mario would seem still to be climbing an invisible vine---or the \emph{map water walk} glitch in \ac{SMB3} where if Mario used a warp whistle while on a boat, he would be able to walk on the sea once he had arrived at the warp island, suggesting a \emph{can move over water} boolean had not been cleaned up.

For \ac{SM64} the biggest source of bugs belonged to \ac{CWE-367}---yet no other game had a single bug assigned to the category.  TOCTOU glitches in \ac{SM64} appear where a game does a check and decides an event ought to start, but then decides to handle other things first:
for example the \emph{MIPS}\footnote{A yellow rabbit named after the N64's processor architecture.} \emph{clips} where players pick up MIPS the rabbit and then place their backs against a wall. The game checks that they are in bounds whilst holding the rabbit, but on placing the rabbit down the game moves Mario a few pixels backwards to account for the model of Mario holding MIPS being bigger than the normal model of Mario.  These few pixels of movement are done without bounds checking and so can be used to clip through doors and access regions of the game normally locked until the player has made sufficient progress---and enabling the game to be beaten more quickly.
Another example of a TOCTOU bug is the \emph{Beat Koopa the Quick in zero seconds} glitch.  If Mario goes near Koopa the Quick the game starts a race to a flag pole.  If Mario is jumping when he is near, then the game waits until Mario lands before starting the race.  By finding a series of cannons that fling Mario through the air and by flying from cannon to cannon, a player can trigger the race start, but not trigger the timer starting until they land.  If they chain the cannons together so that when Mario eventually lands he is at the race flag pole, then the race immediately ends with a time of zero seconds.  The issue is that the whilst the race start event was triggered when Mario was near Koopa the Quick it was not used until much later when the event no longer made any sense and an impossible result occurs

This kingdom of errors also includes two new categories of glitches.  \emph{Failure to Appropriately Recheck Condition} describes when a check should have been made again but is not (for example, if in \ac{SMB3} Mario scrolls a vine off-screen while a vine is growing then the vine stops growing).  \emph{State Exit Too Early} captures when the game changes state before the next state is ready to start---for example in \ac{SM64} if Mario dismounts a Koopa shell immediately upon getting onto it, then the level's music restarts playing from the beginning, despite never playing the Koopa shell-riding music.

\subsection{7PK5: Errors}
\label{sssec:7pk5}

The second-biggest kingdom of errors observed in the Mario games was the \emph{Errors} kingdom: which captures errors in exception handling mechanisms.  Within this kingdom the majority (29 or 63\% of the kingdom) of the glitches were attributed to \ac{CWE-703}.  \Ac{CWE-703} captures what the speedrunning community would call \emph{jank} and what software engineers would call \emph{undefined behavior}.  When the game enters a state which the programmer's did not account for, sometimes strange things happen.  For example in \ac{SM64} if the Yoshi on top of the castle can be persuaded (through extensive manipulation of the random number generation) to step onto a sloped portion of the roof. Then, when the Yoshi is interacted with, it may do something odd including running off the roof, disappearing, or crashing the game by running in a never ending cut-scene.

In addition to \ac{CWE-703}'s undefined behavior, 14 glitches (30\% of the kingdom) were ascribed to a new category, \emph{correct but surprising behavior}, that captures when the game \emph{probably} does the right thing, and yet the behavior is odd and the glitch is notable.   Examples include an \emph{unreachable secret ending} that had been left in a level in \ac{SMB3}, the ability to gain infinitely many coins\footnote{Normally there is a limit of 100.} in \ac{SM64} by destroying \emph{Bowser's Fire}, and a \emph{pipe death} from a scrolling screen if Mario enters a pipe too early in \ac{SMB3}.  In each case the behavior seems reasonable: artifacts from old level designs are no more dangerous than unused code, whilst there may be a limit on coins if there is a mechanism for spawning more you ought to be able to collect them, and is running too far ahead in an auto-scrolling level then it is fair that he should lose a life\ldots{}it is just a bit odd that he loses that life \emph{after} he enters the pipe to the next section.  As such the glitch is correct behavior, but surprising.  We categorized these glitches into the \emph{Errors} kingdom as these bugs are all edge cases where the game does something odd that it probably ought to have an exception handler for, yet it does not.

Finally, three bugs in this kingdom were attributed to a new weakness: \emph{improper correction of illegal state}, an example of which is the \emph{wall moonwalk} in \ac{SMB1}.    If Mario manages to get into a wall (say by momentum or being pushed) then the game attempts to correct this by moving Mario backwards until he is out of the wall.  If Mario turns around just before being pushed into the wall then the game still corrects his position as before, but now pushing Mario through the wall behind him to the other side allowing speedrunners to skip whole sections of the game.

\subsection{7PK1: Input Validation and Representation}

In the \acp{7PK} the \emph{Input Validation and Representation} kingdom (which covers numeric errors, buffer overflows, and pointer errors) is treated as the most important~\cite{tsipenyuk2005seven}, yet only a third of the glitches we saw could be attributed to it.  Whilst half ($\frac{16}{32}$) of the glitches in \ac{SMB1} were in this kingdom, subsequent games suffered less.

Various integer errors (CWE 190--193) affect all the games to varying degrees.  In \ac{SMB1} if Mario loses a single life when he has 128 lives results in a \emph{game over} (the \emph{128 Life Overflow} glitch).  Similar issues affect other Mario games, but first we need to talk about parallel universes.  In \ac{SM64} Mario's position is stored as signed 32-bit floats, but collision checks are done by casting to signed 16-bit integers.  Abusing these checks by manipulating Mario's momentum allows for the creation of \emph{``parallel universes''} caused by integer truncation which can be used to move through levels slowly\footnote{A runner would need to build Mario's momentum for 12 hours through other glitches.} in speedrun categories using limited button presses~\cite{koek2016monty}.

One new weakness was created for this kingdom:  \emph{Entity Limit} glitches relate to issues surrounding which sprites and the number of sprites on the screen at any time and are conceptually similar to some heap-based weaknesses.  If all the entity slots are filled then the game may not be able to load new sprites.  This can lead to invisible enemies (such as the \emph{Invisible Piranha Plant} in level 8--4 of \ac{SMB1} and the \emph{Invisible Muncher} in world 7 of \ac{SMB3}), or despawning power-ups in \ac{SMB1} when the \emph{Power-up Limit} is reached.

\subsection{7PK6: Code Quality}

A relatively small number of glitches (21 or 9\%) could be ascribed to the \emph{Code Quality} kingdom with glitches that seemed to arise from the games' logic either missing checks for certain conditions.  One new category of glitches was ascribed to this kingdom: \emph{incorrectly shared behavior}.  The incorrectly shared behavior category captured when two objects seemed to behave the same, despite being different.  An example of this category is the \emph{Bullet Lift} in \ac{SMB1} where a Bullet Bill will behave as a scale lift (and sometimes leave a trailing string) if the timer is sufficiently low in level 6--3.   

\subsection{7PK*: Environment}
The \emph{Environment} Pernicious Kingdom captures the bugs caused by the interaction of code with everything other than code.  One new weakness was assigned to this kingdom: \emph{Bad Boundary Condition} which captures when in-game objects are poorly arranged so that a player can access areas that should normally be inaccessible.  For example in \ac{SM64} the \emph{Climb the castle without 120 Stars} involves Mario precisely jumping off a boundary wall in order to access a roof that would normally be inaccessible. The \emph{Clip Behind Chan Chomp's Gate} glitch (also \ac{SM64}) a bob-omb is used to provide Mario with enough momentum to push through a gate.  Since each of these glitches are bugs in the definition of the levels rather than the game itself, the category is assigned to the \emph{Environment} kingdom.
Whilst most of this category of glitches were in the 3D~\ac{SM64} ($\frac{16}{20}$) odd examples could be found in the earlier 2D Mario games. For example in \ac{SMW} the \emph{Shell through wall} glitch allows Mario to throw a shell through a steep slope instead of along it.

\subsection{7PK2: API Abuse}
Just one glitch could be attributed to API abuse: \ac{CWE-252}.  The glitch exists in \ac{SM64} and involves \emph{Monty Mole}.  The mole emerges from a series of holes and every time the mole is killed the game runs a function to decide which hole the mole should next emerge.  If the mole is killed by throwing an exploding enemy from a great distance no hole is suitable and the function returns \texttt{NULL}~\cite{koek2016monty}.  This leads to a crash.
Other glitches could be described by this category---glitches assigned to \ac{CWE-703} could be caused by failure to handle a function returning \texttt{NULL}\ldots{}but without analyzing the source code, this sole bug was explained as being an unchecked return value and so mapped to this kingdom.

\subsection{Missing Kingdoms}
Of the original eight \acp{7PK}, we found two kingdoms that were entirely absent in the glitches we examined---\emph{7PK3: Security Features} and \emph{7PK7: Encapsulation}---and one kingdom with a lone glitch attributable to it (\emph{7PK2: API Abuse}).  Why were these kingdoms absent?

The absence of \emph{Security Features} is simple: none of the games have any form of software copy protection or security features.  Instead of copy protection in software the games were distributed on proprietary ROM cartridges, rendering it impossible to those without specialist hardware and knowledge.  This is not to say that games from this era did not have security features and interesting glitches based on it, just that these games did not.  A case study of other games (for instance, the SNES's \emph{Earthbound} had code to increase the difficulty of enemies to superhuman levels if it detected it had been copied) may well find some glitches in this category.

The missing \emph{Encapsulation} and \emph{API Abuse} kingdoms could be explained by the how we analyzed the glitches in each game:
by speculating on the presentation of the bug rather than an analysis of the original source code (which is proprietary).  Because we do not have access to the source code, identifying issues as an API or encapsulation issue is difficult and it is likely some glitches which could have been categorized in these kingdoms were ascribed to another (in particular \ac{CWE-703} is suspect as it captures issues arising from exceptional conditions, rather than the code that led to the exceptional condition in the first place).

\vspace{\baselineskip}\noindent
%  \textbf{Summary:} We find that \input{data/out/glitch-total}glitches in the 4 Mario games can be categorized using 6 out of 8 Pernicious Kingdoms, and mapped to \input{data/out/codes-total}weaknesses---including \input{data/out/new-codes-total}new weaknesses.  The majority of the glitches can be described as \emph{Time and State} issues.
  \textbf{Summary:} Glitches that speedrunners exploit map to conventional software weakness taxonomies like \ac{CWE}.  There appear to be 7kinds of glitch that do not have counterparts in \ac{CWE} but which do fall into the broader Pernicious Kingdoms.

\section{Discussion}

\subsection{Growing expertise}

The examples we studied represent some of the earliest videogames created by Nintendo.   How did the bugs change as the games grew in complexity, and Nintendo's experience programming them grew?

%\begin{figure}
%  \tablestyle
%  \includegraphics[width=\linewidth]{figures/out/vs-kingdoms.pdf}
%  \caption{Changing percentages of glitch categories over time in Super Mario Games.}
%  \label{fig:vs-kingdoms}
%\end{figure}

For 1985's \ac{SMB1}, 50\% of the glitches we studied relate to \emph{7PK1: Input Validation and Representation}.  By 1996 and \ac{SM64} this kingdom accounts for just 5\% of the glitches.
In contrast, for \ac{SMB1} 28\% of the glitches could be attributed to \emph{7PK4: Time and State}, but by \ac{SM64} over half (54\%) the glitches relate to it: why the change?
Perhaps it is experience.  Toshihiko Nakago was the programmer for \ac{SMB1}, \ac{SMB3} and \ac{SMW}; and the \emph{Input Validation and Representation} kingdom mostly basic programming errors.  Both \ac{SMB1} and \ac{SMB3} would have been written in assembly and by the time of \ac{SMW} and \ac{SM64}, which would have both been written in C, improved diagnostics, a higher level language, and the increased experience meant simpler errors were less likely to be made, hence fewer of those sorts of glitches.

In contrast, the glitches attributed to the \emph{Time and State} kingdom increased as the games became more complex.  Whilst \ac{SMB1} was implemented in just over 8,000 lines of assembly\footnote{Excluding data, comments and labels: \url{https://gist.github.com/1wErt3r/4048722}}; a reproduction of \ac{SM64}'s source code\footnote{\url{https://github.com/n64decomp/sm64}} includes over 600,000 lines of code (the majority in C).  More code, more state and more complex interactions
% \footnote{In \ac{SMB1} Mario can interact with enemies by jumping on their head, but by \ac{SM64} Mario can talk, chase and throw them as well as still jumping on their heads.}
meant more glitches related to the increased amount of state.  

Nintendo still makes Super Mario games: are they still having these issues almost 40 years later?  An informal comparison can be made to 2019's \emph{Super Mario Maker 2} which lets players create and share levels made in the style of \ac{SMB1}, \ac{SMB3} and \ac{SMW}---despite not sharing code with the original games.  A quick survey of the glitches listed on the Super Mario Wiki suggest glitches to clip through surfaces, accelerating enemies, and surrounding the interactions between Mario, enemies and slopes persist---just as it was in 1985.

\subsection{Game specific weaknesses?}
In our analysis of the 237glitches we found 7categories of glitch that did not seem to have a code in \ac{CWE} (\cref{tab:codebook}).  This raises the question: are the issues we identified so specific to games that they have no counterpart in more general software?

A few of the glitches we identified are analogous to bugs in conventional software.  \emph{Correct but surprising behavior} captures when software does what it should but nevertheless surprises users---this seems similar to a usability issue where software works as the programmer intended but its users are surprised by how to use it.  The \ac{CWE} taxonomy for the most part does not include usability issues but perhaps they should be. Issues with how a game is made have led odd behavior---this should be tracked the same as any correctness bug.  A counterargument could be made that usability issues represent issues with the user's mental models or documentation rather than the code itself, but it is somewhat unsatisfying: bugs get onto lists of glitches by exhibiting odd behavior in games.

The \emph{Incorrectly Shared Behavior} weakness could be seen as an object-oriented programming error where two objects mistakenly inherit from a shared base class.  \emph{Failure to Appropriately Recheck Condition} and \emph{State Exit Too Early} seem similar to control flow issues---though perhaps more specific to games or state machines. \emph{Improper Correction of Illegal State} could be a child of \ac{CWE-703} (or \ac{CWE-755}), yet the examples given in \ac{CWE} focus on the handling of \emph{an exception}, rather than the more user-forward explanation that the game is behaving weirdly having tried to fix itself from being in another weird state.  It is conceivable that all these weaknesses exist in most other forms of software.

Most software does not have \emph{Entity Limits} in the same way that many games do (a means to avoid spawning too many items at once, slowing down the game).  But general software frequently does have to work with limited resources---be that pools of threads or heaps of memory.  \Ac{CWE-400} and \ac{CWE-770} are two \acp{CWE} that capture weaknesses coming from limited resources. Yet the consequences of these issues is different in other software than in games.  In games running out of entity slots leads to parts of levels going missing---allowing speedrunners to bypass in game barriers or avoid enemies---whereas in more general software reaching such limits results in throttling and resource exhaustion.

The \emph{Bad Boundary Definition} weakness is unlikely to be seen in other software: most software does not involve moving an actor around an environment with boundaries to begin with, but might this change?  Virtual Reality systems are increasingly normal.  As general purpose user interfaces adapt to these new ways of interacting computers, will we see other developers make the same mistakes that game programmers have been made decades?

%\vspace{\baselineskip}\noindent
More importantly do the new weaknesses we have identified generalize?  That is do these glitches occur in more games than just the 4 Super Mario games we studied?

\subsection{Comparisons With Other Taxonomies}

To try and answer if these glitches are more general we can compare with other work on game glitches.
Our work is not the first to study glitches in games, but it is the first to make links between glitches and conventional software weaknesses.  How do the \acp{CWE} we observeed relate to earlier glitch taxonomies?
Bainbridge and Bainbridge's \emph{Model  Discrepancy}~\cite{bainbridge2007creative} and Lewis~et~al{.}'s \emph{Invalid Position} seem to be capturing a similar range of issues as our \emph{Bad Boundary Definition} weakness, further suggesting that it is a game-specific issue that occurs accross \emph{all} games and not just Super Mario games.

The other code that seems universal to all 3 attempts to categorize glitches is what Bainbridge and Bainbridge called \emph{Bizarre Maneuver}, Lewis~et~al{.} called \emph{Artificial Stupidity} and what we categorized as either \ac{CWE-703} or \emph{Correct But Surprising Behavior}.   In all cases these are glitches where the game made a decision about how to handle a weird edge case, caused in Bainbridge and Bainbridge's taxonomy by the player entering a bizarre series of inputs, in Lewis~et~al{.}'s case by the game making a \emph{stupid} decision how to handle an edge case, and by doing something correct but odd, or just plain wrong in our taxonomy.   Like the bad boundary definitions, these issues seem to be a game-specific problem caused in part by the need for games to keep running in the face of unusual player actions, rather than denying actions or crashing as you would in other software.  From the player's perspective this tradeoff makes a lot of sense---better for a game to keep going and make an attempt to fix weirdness than crash and destroy a player's progress---but it also leads to interesting glitches which speedrunners exploit to go fast.  By ensuring that games remain crash free, the games also enable weirdness in some edge cases.  Perhaps, given that the behavior is such an integral part of speedrunning, instead of attempting to name the weakness or link it to \ac{CWE} we ought to use the speedrunning community's name for this and just say that some software (and games in particular) are just a bit \emph{janky}?

\section{Conclusions}

In a case study of four Mario games and  237glitches we find bugs that are exploited by speedrunners to let them to beat games quickly.  We find 7new weaknesses missed by \ac{CWE}.

Should these new weakness enter \ac{CWE}? It depends on what the purpose of tracking them is and whether they generalize.  \ac{CWE} is a large taxonomy, but its job is to serve as a common language for talking about weaknesses in software. Glitches that form a category of weakness identified in multiple studies and without representation in \ac{CWE} (such as \emph{Bad Boundary Definition}) probably should be included.  The \emph{Correct But Surprising}  where there is a sensible analogy to be made with software usability are also arguable.
The counter argment, however, is that glitches in games, traditionally, were seldom patched (though with online distribution of games this is changing).  Whilst with other software we attempt to remove all the bugs, for games a limited amount of jank is desirable because it is part of what makes the games fun~\cite{marshall2021janky}.

The point of \ac{CWE} is to give a common language to help aid with \emph{fixing} bugs.  In contrast the point of speedrunning is as an art built around to \emph{appreciating} the glitches that let us beat games fast.  Whilst speedrunning is not a software engineering discipline the tricks they employ do come in part from software engineering flaws.
As engineers we can appreciate the performance and skill of speedrunners in breaking our code, but perhaps we do not need a standard language to fix these issues---unless they also appear in other software.
Perhaps some software is best left broken, and perhaps sometimes its fun to watch Mario fly through a wall and enjoy Super Mario for what it is: fun.

%\paragraph{Data Availability}
%\noindent
%All data and analysis  is available online at: {\ttfamily \url{http://redacted.github.repo}}.

\bibliographystyle{ACM-Reference-Format}
\balance
\bibliography{/home/joseph/Notes/bibliography,other}

\end{document}